\documentclass[journal]{IEEEtran}
\usepackage{amssymb}

\usepackage{amsfonts}
\usepackage{bbm}
\usepackage{amsfonts}
\usepackage[dvips]{graphicx}
\usepackage[dvips]{color}
\usepackage{graphicx}
\usepackage{amsmath}
\usepackage{fancyhdr}
\usepackage{stfloats}
\usepackage{multirow}
\usepackage{algorithm}
\usepackage{algorithmic}
\usepackage{cite}
\usepackage[bookmarks=false]{}
\usepackage{amsthm}
\usepackage{algorithm}
\usepackage{algorithmic}
\usepackage{mathbbol}
\usepackage{amsfonts}
\usepackage{graphicx}
\usepackage{amsmath}
\usepackage{amssymb}
\usepackage{latexsym}
\usepackage{graphicx}
\usepackage{cite}
\usepackage{url}
\usepackage{stfloats}
\usepackage{mathrsfs}
\usepackage{setspace}
\usepackage{booktabs}
\usepackage{latexsym}
\usepackage{url}
\usepackage{stfloats}
\usepackage{mathrsfs}
\theoremstyle{plain}

\usepackage{cite}
\usepackage{setspace}
\usepackage{textcomp}
\usepackage{xcolor}
\usepackage{makecell}
\usepackage{etoolbox}
\usepackage{diagbox}
\usepackage{caption}
\usepackage{amsthm}
\usepackage{multirow}
\usepackage{array}
\usepackage{colortbl}
\usepackage{bbding}
\usepackage[tight,footnotesize]{subfigure}
\usepackage[]{hyperref}
\hypersetup{
	bookmarksnumbered=true,     
	bookmarksopen=true,         
	bookmarksopenlevel=1,       
	colorlinks=true,
	linkcolor=black            
}

\begin{document}
\clearpage
\title{\huge OFDM-Based Digital Semantic Communication with Importance Awareness}
%


\author{Chuanhong Liu
, Caili Guo, \emph{Senior Member}, \emph{IEEE}, Yang Yang, Wanli Ni, and Tony Q.S. Quek, \emph{Fellow}, \emph{IEEE}
\thanks{This work was supported in part by the Fundamental Research Funds for the Central Universities (No.2021XD-A01-1) and in part by BUPT Excellent Ph.D. Students Foundation (No.CX2022101).}
\thanks{Chuanhong Liu and Caili Guo are with the Beijing Key Laboratory of Network System Architecture and Convergence, School of Information and Communication Engineering, Beijing University of Posts and Telecommunications, Beijing 100876, China (e-mail:2016\_liuchuanhong@bupt.edu.cn).}
\thanks{Yang Yang is with the Beijing Laboratory of Advanced Information Networks, School of Information and Communication Engineering, Beijing University of Posts and Telecommunications, Beijing 100876, China (e-mail:yangyang01@bupt.edu.cn).}
\thanks{Wanli Ni is with Department of Electronic Engineering, Tsinghua University, Beijing 100084, China (e-mail: charleswall@bupt.edu.cn).}
\thanks{Tony Q. S. Quek is with the Dept. of Information Systems Technology and Design, Singapore University of Technology and Design, Singapore, 487372 (e-mail: tonyquek@sutd.edu.sg).}
}
\maketitle
\pagestyle{headings}
\vspace{0cm}
\begin{abstract}
Semantic communication (SemCom) has received considerable attention for its ability to reduce data transmission size while maintaining task performance. However, existing works mainly focus on analog SemCom with simple channel models, which may limit its practical application. To reduce this gap, we propose an orthogonal frequency division multiplexing (OFDM)-based SemCom system that is compatible with existing digital communication infrastructures. In the considered system, the extracted semantics is quantized by scalar quantizers, transformed into OFDM signal, and then transmitted over the frequency-selective channel. Moreover, we propose a semantic importance measurement method to build the relationship between target task and semantic features. Based on semantic importance, we formulate a sub-carrier and bit allocation problem to maximize communication performance. However, the optimization objective function cannot be accurately characterized using a mathematical expression due to the neural network-based semantic codec. Given the complex nature of the problem, we first propose a low-complexity sub-carrier allocation method that assigns sub-carriers with better channel conditions to more critical semantics. Then, we propose a deep reinforcement learning-based bit allocation algorithm with dynamic action space. Simulation results demonstrate that the proposed system achieves 9.7\% and 28.7\% performance gains compared to analog SemCom and conventional bit-based communication systems, respectively.
\end{abstract}

\begin{IEEEkeywords}
	Semantic communication, OFDM, bit allocation, reinforcement learning, semantic importance.
\end{IEEEkeywords}

\vspace{-0.3cm}
\section{Introduction}
\label{sec:intro}

\IEEEPARstart{U}{nder} the umbrella of 6G network support, the inexorable trend revolves around the deep integration of Artificial Intelligence (AI) and wireless communication networks. As intelligent devices interconnect on a massive scale and wireless data traffic experiences a tremendous surge, the looming challenge of spectrum scarcity has come to the forefront, presenting substantial hurdles for 6G wireless communication\cite{zhang-DRL}. However, existing communication technologies primarily prioritize the precise transmission of each symbol, often neglecting the ultimate task and the inherent meaning carried within the transmitted data. This oversight results in the unnecessary depletion of wireless communication resources and falls short of addressing the burgeoning demands of future large-scale communications. Semantic communication (SemCom)\cite{Farshbafan-Curriculum}, a technique capable of notably enhancing communication efficiency and robustness, is now acknowledged as a foundational technology driving the progression of sixth-generation wireless networks\cite{framework_Yang}.

In contrast to traditional communication systems that transmit all bits indiscriminately, SemCom focuses on extracting and transmitting the most valuable task-oriented semantics from the source data\cite{Shao_IB}. By doing so, SemCom can bring several benefits. First, it mitigates a substantial amount of redundant or unnecessary semantic overhead, thereby significantly enhancing the communication efficiency and addressing the challenge of spectrum scarcity. Second, it enhances the robustness of the communication system against channel noise by learning the characteristics of the wireless channel. Third, it achieves better performance of messages recovering and task completion with the help of knowledge bases. In addition, the ever-increasing deployment of wireless devices, placing unprecedented requirements on spectrum and energy, underscores the need for a more in-depth exploration of SemCom.

\vspace{-0.2cm}
\subsection{Related Works}
\textit{1) Analog semantic communication}: In analog SemCom, a notable contribution is the work by Xie et al. \cite{Xie_Deep}, where they introduced a robust Transformer-based SemCom system for text transmission. This system demonstrates resilience to channel variations and exhibits applicability across diverse scenarios. The work in \cite{Gunduz_JSCC} presented a joint source-channel coding scheme based on convolutional neural networks (CNN) to transmit image data in a wireless channel, which can jointly optimize the source-channel encoder and decoder. The authors in \cite{MU-DeepSC1} considered a multi-modal transmission scenario and proposed a multi-users SemCom system for serving visual question-answering tasks, in which Long Short Term Memory is used for text transmitter and CNN for image transmitter. In all of the aforementioned works, semantic features are directly transmitted or mapped into analog symbols for transmission. However, analog SemCom does not match up with real-world digital systems. Furthermore, digital communication offers numerous advantages over analog communication, including higher error correction capability, stronger anti-interference ability, etc.

\textit{2) Digital semantic communication}: To make SemCom compatible with the existing digital communication system, some researchers have devoted efforts to developing digital SemCom systems \cite{Hu_VQ, Nemati_VQ, Fu_VQ, Bo_VQ, Guo_DSC, Huang_Toward}. The authors in \cite{Hu_VQ} designed a robust vector quantized SemCom system to mitigate semantic noise, which outperforms traditional methods on a classification task. The work in \cite{Nemati_VQ} proposed a vector quantized DL-based JSCC system and showed its robustness against noisy wireless channels. The authors in \cite{Fu_VQ} proposed a vector quantization-based SemCom system specifically designed for image transmission. The authors in \cite{Bo_VQ} developed a joint coding-modulation scheme for digital SemComs with BPSK modulation. Notably, all of the mentioned works \cite{Hu_VQ, Nemati_VQ, Fu_VQ, Bo_VQ} incorporate vector quantization as a key component of their respective approaches. However, quantizers are implemented in digital signal processing systems using analog-to-digital converters (ADCs), which typically operate in a serial scalar manner due to hardware-limitations \cite{Neil_hard}. The work in \cite{Guo_DSC} proposed a novel non-linear quantization module with trainable quantization levels to efficiently quantifies semantic features. The authors in \cite{Huang_Toward} designed a novel reinforcement learning (RL)-based semantic bit allocation model with adaptive quantization levels. The aforementioned works provide some insights and remarks for the design of digital SemCom. However, there are still some key issues remaining unexplored.

\vspace{-0.2cm}
\subsection{Challenges and Contributions}
Deep learning-based end-to-end (E2E) communication system relies on a differentiable channel model to facilitate the learning process\cite{E2Echannel}. However, complications arise when the channel parameters are not pre-known, as the gradients become obstructed by the unknown channel during back-propagation\cite{Qin_physical}. Consequently, existing research has predominantly concentrated on simple channel models, such as additive white Gaussian noise (AWGN) channel. Regrettably, this approach tends to overlook the ramifications of intricate channel models, like the multipath channel, which introduces challenges like inter-symbol interference (ISI) capable of significantly deteriorating the SemCom performance. Fortunately, the profound integration of SemCom and digital communication technologies may offer a promising solution to this issue. 

In addition, semantics possess non-uniform characteristics, signifying that different semantics hold varying semantic importance in accomplishing intelligent tasks. The study in \cite{Chen_allocation} has demonstrated the effectiveness of importance-aware resource allocation in enhancing both resource utilization and task performance, which demonstrates the necessity of considering semantic importance. However, the effective evaluation of task-oriented semantic importance requires further research. Moreover, to fully harness the potential benefits of digital SemCom, there is a pressing need to develop more efficient resource allocation strategies including both bit and sub-carrier allocation. These strategies should allocate the finite communication resources in a task-oriented fashion, prioritizing data with more important semantic information. When delving into performance optimization for digital SemCom with importance awareness, we encounter the following challenges:

\begin{itemize}
\item[$\bullet$] \emph{Challenge 1: How to develop a digital SemCom that is resilient to multipath fading and can seamlessly integrate with existing digital communication infrastructure?}

\item[$\bullet$]  \emph{Challenge 2: How to measure the semantic importance given a specific task?}

\item[$\bullet$]  \emph{Challenge 3: How to effectively allocate the sub-carriers and bits to different semantics based on their respective importance?}
\end{itemize}

In this paper, we first develop an orthogonal frequency division multiplexing (OFDM)-based digital SemCom framework. Then, we propose a novel semantic importance measurement method and devise a deep reinforcement learning-based algorithm for semantic importance-aware sub-carrier and bit allocation. To our best knowledge, \emph{this is the first work that studies ODFM-based digital SemCom considering the semantic importance.} The main contributions of this paper are summarized as follows:

\begin{itemize}
    \item[1)] We propose an OFDM-based digital SemCom framework that employs scalar quantization for hardware-limited scenarios. This framework seamlessly integrates with existing digital communication systems and can effectively mitigate the impact of multipath fading. Considering the varying importance of different semantics, we formulate a sub-carrier and bit allocation problem, with the objective of minimizing semantic distortion while simultaneously maximizing task performance. This addresses the aforementioned \emph{Challenge 1}.
    \item[2)] To solve the problem, we first introduce a semantic importance measurement method that considers both the correlation between semantics and tasks and the correlation between different semantics. The former is derived by computing the gradient of semantic features using task results, while the latter is obtained through cosine similarity analysis. By combining these two factors, we are able to comprehensively evaluate the semantic importance. This addresses the aforementioned \emph{Challenge 2}.
	\item[3)] Then, we propose a low-complexity sub-carrier allocation method and a reinforcement learning-based bit allocation algorithm. Specifically, the algorithm incorporates both semantic distortion and task performance into the reward function, allowing for a comprehensive evaluation of the communication process. Furthermore, a dynamic action space is designed to adjust the available choices of bit allocation according to the number of allocated bits. This dynamic action space empowers the system to explore and select diverse quantization strategies, thereby enhancing overall performance optimization. This addresses the aforementioned \emph{Challenge 3}.
    \item[4)] Experimental results demonstrate that the proposed digital SemCom outperform both analog SemCom and conventional communication system in terms of compression and task performance. Additionally, the proposed bit allocation algorithm yields substantial improvements in task performance across different signal-to-noise ratio (SNR) regimes and varying bit constraints.
\end{itemize}

The remainder of this paper is organized as follows. The system model and problem formulation are described in Section II. Section III introduces the proposed semantic importance measurement method. The sub-carrier and RL-based bit allocation algorithm is presented in Section IV.  Experiments results are analyzed in Section V. Finally, Section VI draws critical conclusions.


\section{System Model and Problem Formulation}
\label{sec:system}

\begin{figure*}[t]
	\begin{center}
		\includegraphics[width=0.8\linewidth]{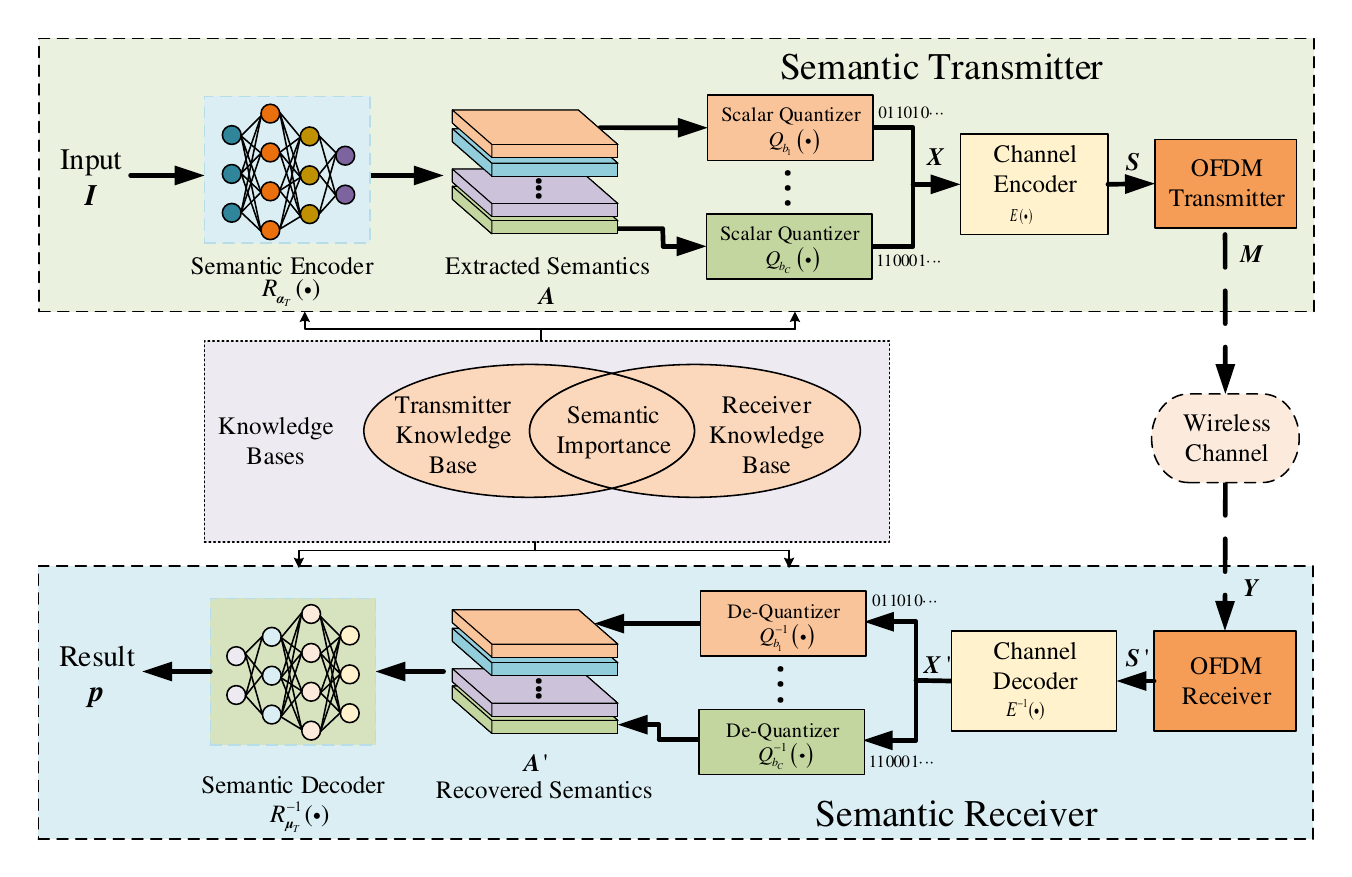}
	\end{center}
	\caption{System model of OFDM-based digital semantic communication.}
	\label{fig:model}
\end{figure*}

\subsection{System Model}
As shown in Fig. \ref{fig:model}, we consider an E2E OFDM-based digital SemCom system constructed by a neural network architecture. Specifically, the transmitter consists of a semantic encoder to extract the semantic features from the source data, a series of scalar quantizers to quantize the continuous semantics to digital bits based on the allocated quantization bits, a channel encoder to combat interference, and an OFDM transmitter to generate OFDM symbols to facilitate the transmission subsequently. The receiver is composited with an OFDM receiver for symbol detection, a channel decoder, a series of de-quantizers for semantics recovery, and a semantic decoder for task execution. Both the user and BS have access to a knowledge base, which can guide the semantic extraction and task execution processes.

At the semantic transmitter, neural networks are first utilized to extract the semantic information from source data $\boldsymbol{I}$, which can be denoted by
\begin{eqnarray}\label{signalform}
	{\boldsymbol {A}} = {R_{\boldsymbol{\alpha}_T}}({\boldsymbol {I}}),
\end{eqnarray}
where ${R_{\boldsymbol {\alpha}_T}}(\cdot)$ denotes the semantic encoder with the parameter set $\boldsymbol{\alpha}_T$, and ${{T}}$ is the intelligent task. The output ${\boldsymbol{A}}{\rm{ = [}}{{\boldsymbol{a}}_{\rm{1}}}{\rm{,}}...{\rm{,}}{{\boldsymbol{a}}_{{i}}}{\rm{,}}...{\rm{,}}{{\boldsymbol{a}}_{{C}}}{\rm{]}}$ is a series of semantic features, where $C$ is the number of semantics. The elements in ${\boldsymbol {A}}$ are continuous and cannot be directly transmitted via digital communication system. Given a specific task, different semantic features make varying contributions to the accomplishment of intelligent tasks, thereby possessing different semantic importance. Here, the importance weights of semantic information is denoted as ${\boldsymbol{\omega }} = \left[ {{\omega _1},...,{\omega _i},...,{\omega _C}} \right]$. In Sec. \ref{sec:importance}, we will propose a semantic importance measurement method to obtain ${\boldsymbol{\omega }}$.

Then, the semantic features are quantized via a series of scalar quantizers to transform them into a bitstream, which can be denoted by
\begin{eqnarray}\label{}
	{\boldsymbol{x_i}} = {Q_{b_i} }({\boldsymbol{a_i}}), i \in {1,2,...,C}
\end{eqnarray}
where ${Q_{b_i} }(\cdot)$ denotes the $i$-th scalar quantizer, $b_i$ is the number of quantization bits, and $\boldsymbol b=[b_1,...,b_i,...b_C]$ is the bit allocation vector. The number of scalar quantizers corresponds to the number of semantics and is denoted as $C$. The quantization results can be represented as ${\boldsymbol{X}}{\rm{ = [}}{{\boldsymbol{x}}_{\rm{1}}}{\rm{,}}...{\rm{,}}{{\boldsymbol{x}}_{{i}}}{\rm{,}}...{\rm{,}}{{\boldsymbol{x}}_{{C}}}{\rm{]}}$. In particular, the scalar quantizers ${Q_{b_i} }(\cdot)$ implement non-subtractive uniform dithered quantization\cite{Gray_quantizer}. Unlike subtractive dithered quantization, non-subtractive quantizers do not require the realization of the dithered signal to be subtracted from the quantizer output in the digital domain, resulting in a practical structure\cite{Gray_quantizer}. Let $\gamma$ denote the dynamic range of the quantizer, and define ${\Delta _i} = \frac{{2\gamma }}{{{M_i}}}$ as the quantization spacing, where ${M_i} = {2^{{b_i}}}$ is the number of quantization levels. Therefore, the output of the scalar quantizers with input $y$ can be written as ${Q_{b_i} }(y) = q_i(y + z)$, where $z$ is i.i.d. random variable uniformly distributed over $\left[ { - \frac{{{\Delta _i}}}{2},\frac{{{\Delta _i}}}{2}} \right]$, independent of the input, representing the dither signal. The function ${q_i}(\cdot)$, which implements the uniform quantization, is given by
\begin{eqnarray}\label{}
{q_i}(y) = \left\{ {\begin{array}{*{20}{c}}
{ - \gamma  + {\Delta _i}(l + 0.5),\begin{array}{*{20}{c}}
{y - l{\Delta _i} + \gamma  \in [0,{\Delta _i}]}\\
{l \in \{ 0,1,...,M_i - 1\} }
\end{array}}\\
{\rm{sign}(y)(\gamma  - \frac{{{\Delta _i}}}{2}),\left| y \right| > \gamma }
\end{array}} \right.
\end{eqnarray}

Note that when $b_i=1$, the corresponding quantizer is a standard one-bit sign quantizer of the form $q_i(y)=c\cdot \rm{sign}(y)$, where the constant $c>0$ is determined by the dynamic range.

Next, the quantized bitstream is encoded by a channel encoder to combat channel noise, which can be denoted by
\begin{eqnarray}\label{}
	{\boldsymbol{S}} = {E}({\boldsymbol{X}}),
\end{eqnarray}
where $E(\cdot)$ denotes the channel encoder, which can be implemented via neural networks and traditional methods (e.g. Low-density parity check (LDPC), Polar, etc.).

Subsequently, the bitstream $\boldsymbol{S}$ is input into the OFDM transmitter and modulated onto $K$ subcarriers, denoted as $[s_1, ..., s_K]$. These OFDM symbols are then transformed into an OFDM signal using an inverse fast Fourier transform (IFFT), which distinguish OFDM from single carrier systems. To address issues like multipath channels and inter-symbol interference, a cyclic prefix (CP) is inserted into the OFDM signal. The CP ensures subcarrier orthogonality and facilitates symbol separation at the receiver. Here, we assume that the pilot symbols are identical without loss of generality. Following these operations, the OFDM signal, denoted as $\boldsymbol{M}$, is transmitted through the multipath channel, encountering propagation effects like fading, noise, and interference.

At the semantic receiver, the received signal can be expressed as
\begin{eqnarray}\label{}
{\boldsymbol{Y}} = {\boldsymbol h}{\boldsymbol{M}} + \boldsymbol{n},
\end{eqnarray}
where ${\boldsymbol h}$ is the channel response in the frequency domain and $\boldsymbol{n}$ is the independent and identically distributed complex Gaussian noise sampled from ${{\cal C}{\cal N}}(0,{\boldsymbol{\sigma }}^2{\boldsymbol I})$.

We assume that each sub-carrier has a bandwidth that is much smaller than the coherence bandwidth of the channel and the instantaneous channel estimations on all the sub-carriers can be estimated at the receiver and transmitted to the transmitter. Consequently, for a frequency-selective channel, $\boldsymbol{h}=[h_1,h_2,...,h_K]$ can be viewed as $K$ flat sub-channels, and each sub-channel has different channel gains \cite{Jiang_video}. Therefore, the demodulated OFDM signal can be estimated by
\begin{eqnarray}\label{}
{\boldsymbol{S}}' = \left[ {\frac{{{y_1}}}{{{{\hat h}_1}}},\frac{{{y_2}}}{{{{\hat h}_2}}},...,\frac{{{y_K}}}{{{{\hat h}_K}}}} \right],
\end{eqnarray}
where ${\boldsymbol{\hat h}} = \left[ {{{\hat h}_1},{{\hat h}_2},...,{{\hat h}_K}} \right]$ is channel state estimation.


The received symbols are decoded to recover bitstream via the channel decoder, which can be expressed as
\begin{eqnarray}\label{}
	{{{\boldsymbol{X}}'}} = {E^{ - 1}}({{\boldsymbol{S}}'}),
\end{eqnarray}
where ${E^{ - 1}}(\cdot)$ denotes the channel decoder.

Then, the bitstream will be dequantized to recover the semantics, which can be represented by
\begin{eqnarray}\label{}
	{{{\boldsymbol{A}}'}} = {Q_{M_i}^{ - 1}}({\boldsymbol{X}}'),
\end{eqnarray}
where ${Q_{M_i}^{ - 1}}(\cdot)$ denotes the scalar de-quantizer, with the quantization level corresponding to the $i$-th scalar quantizer.

Finally, the semantic receiver inputs the recovered semantics ${{\boldsymbol{A}}'}$ into the semantic decoder to complete the intelligent tasks. Specifically, the output can be represneted as
\begin{eqnarray}\label{}
	{\boldsymbol{p}} = {R_{\boldsymbol{\mu_T}}^{ - 1}}({\boldsymbol{{{\boldsymbol{A}}'}}}),
\end{eqnarray}
where ${\boldsymbol{p}}$ is the task result, which will be returned to the transmitter and ${R_{\boldsymbol{\mu}_T}^{ - 1}}(\cdot)$ denotes the semantic decoder with the parameter set ${\boldsymbol{\mu}_T}$. 

The quality of SemCom can be evaluated based on two aspects: semantic distortion and task performance\cite{Shao_IB}, both of which are influenced by bit allocation strategy ${\boldsymbol b}$ and sub-carrier allocation policy ${\boldsymbol{\rho }}$. Considering the semantic importance, the weighted semantic distortion between transmitted semantics $\boldsymbol{A}$ and the received semantics $\boldsymbol{A}'$ is primarily caused by quantization and channel noise, denoted as $d\left( {{\boldsymbol{A}},{\boldsymbol{A}}';{\boldsymbol{b}},{\boldsymbol{\omega }}, {\boldsymbol{\rho }}} \right)$. Specifically, in this work, the distortion function $d(\cdot)$ adopts the extensively used mean square error (MSE) function, which can be computed as
\begin{eqnarray}\label{}
	d\left( {{\boldsymbol{A}},{\boldsymbol{A}}';{\boldsymbol{b}},{\boldsymbol{\omega }},{\boldsymbol{\rho }}} \right) = \sum\limits_{i = 1}^C {\sum\limits_{k = 1}^K {{\rho _{ik}}{\omega _i}{{\left| {{{\boldsymbol{a}}_i} - {\boldsymbol{a}}_i^{'}} \right|}^2}} },
\end{eqnarray}
where ${\omega _i}$ is the semantic importance weight of the $i$-th semantic feature. ${\rho _{ik}} = 1$ means that the $i$-th semantics $\boldsymbol{a}_i$ is transmitted via the $k$-th sub-carrier. On the other hand, the metric for task performance $L(\boldsymbol{p}, \boldsymbol{y}; {\boldsymbol b}, {\boldsymbol{\rho }})$ may vary depending on the specific task, where $\boldsymbol{y}$ is ground truth. For instance, top-1 accuracy is commonly used for classification tasks, whereas mean average precision is employed for detection tasks.

\subsection{Problem Formulation}
Given the defined system model, our objective is to minimize the weighted semantic distortion and simultaneously maximize the task performance by optimizing sub-carrier and bit allocation strategy under the maximum bit constraint. The optimization problem can be formulated as follows

\begin{align}\label{Q1}
&\mathop {\max }\limits_{{\boldsymbol{b}},{\boldsymbol{\rho }}} L({\boldsymbol{p}},{\boldsymbol{y}};{\boldsymbol{b}},{\boldsymbol{\rho }}) - \beta d\left( {{\boldsymbol{A}},{\boldsymbol{A}}';{\boldsymbol{b}},{\boldsymbol{\omega }},{\boldsymbol{\rho }}} \right)\\
&\rm{s.t.}\;\; \sum\limits_{i = 1}^C {{b_i}}  \le B,  \tag{\theequation a}\\
&\;\;\;\;\;\;\; b_i \in \mathbb{N}^+, \;\;\; i = 1,2,...,C \tag{\theequation b}\\
&\;\;\;\;\;\;\; {\rho _{ik}} \in \left\{ {0,1} \right\}, \;\;\; i = 1,2,...,C, \;\; k = 1,2,...,K  \tag{\theequation c}\\
&\;\;\;\;\;\;\; \sum\limits_{i = 1}^C {{\rho _{ik}}}  = 1, \;\;\; k = 1,2,...,K \tag{\theequation d}\\
&\;\;\;\;\;\;\; \sum\limits_{k = 1}^K {{\rho _{ik}}}  = 1, \;\;\; i = 1,2,...,C \tag{\theequation e}
\end{align}
where ${\boldsymbol{\rho }} = \left\{ {{\rho _{ik}}} \right\},i \in \left[ {1,C} \right],k \in \left[ {1,K} \right]$ is the sub-carrier allocation matrix and $\beta$ is a hyper-parameter. Constraint (\ref{Q1}a) indicates that the sum of allocated bits cannot exceed the overall bits. Constraint (\ref{Q1}b) illustrates that the allocated bits can only be positive integer, which is discrete. Constraint (\ref{Q1}c) and (\ref{Q1}d) guarantee that a sub-carrier can be allocated to transmit only one semantics at a time. Constraint (\ref{Q1}e) ensures that each semantics can be transmitted using only one sub-carrier at a time. 

Problem (\ref{Q1}) is challenging to solve by traditional algorithms such as greedy algorithms due to the following reasons. First, the objective function in problem (\ref{Q1}) depends not only on the bit allocation strategy but also on the neural network-based semantic encoder and semantic decoder. Second, the relationship between the non-convex objective function (i.e., the semantic distortion and the task performance) and the optimization variables (i.e., sub-carrier allocation and bit allocation strategy) cannot be accurately characterized using a mathematical expression. To solve problem (\ref{Q1}), we first introduce an efficient and low-complexity method for sub-carrier allocation based on semantic importance. Then, we further propose a reinforcement learning-based algorithm that enables the user to optimize the bit allocation strategy based on the semantic importance to improve the performance of the whole system. Before introducing the proposed algorithm, we first propose a novel semantic importance measurement method to measure the non-equal properties of semantics, which forms the foundation of this work.

\begin{figure*}[t]
	\begin{center}
		\includegraphics[width=0.8\linewidth]{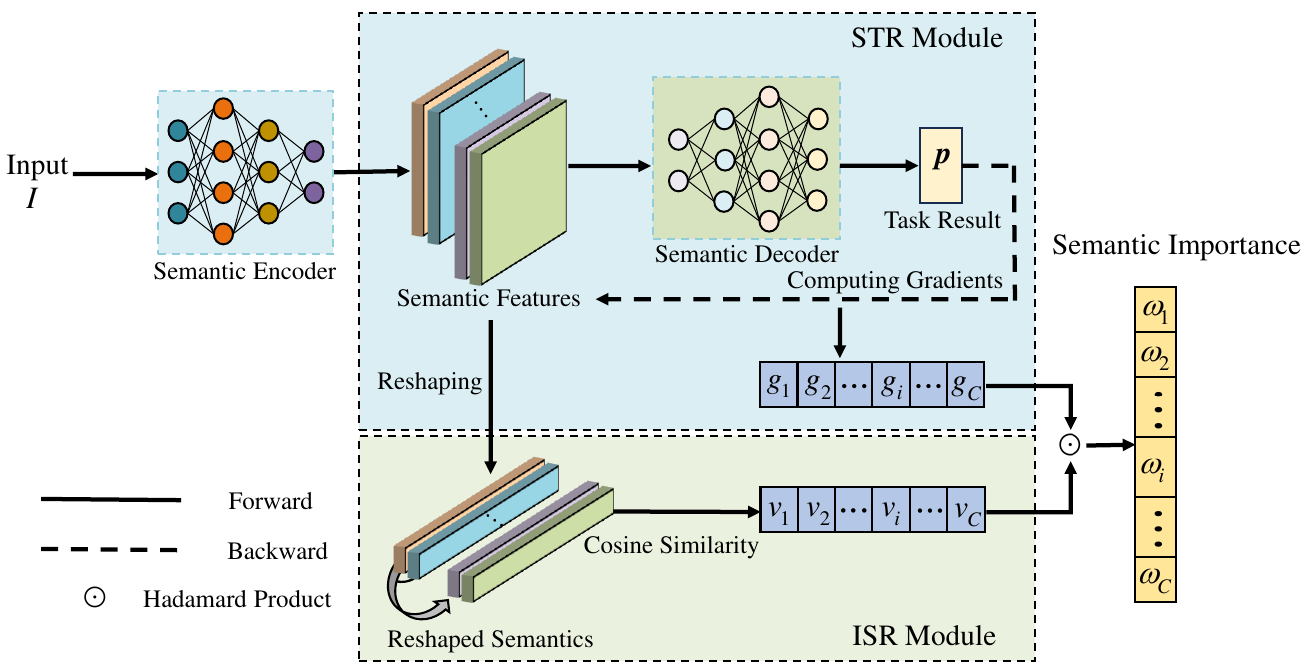}
	\end{center}
	\caption{Illustration of the proposed semantic importance evaluation method.}
	\label{fig:semantic_importance}
\end{figure*}

\section{Semantic Importance Evaluation}
\label{sec:importance}
As shown in Fig. \ref{fig:semantic_importance}, the proposed semantic importance evaluation method primarily comprises two modules: the Semantics Task Relevance (STR) module and the Inter-Semantics Relevance (ISR) module. Building on the insights from \cite{framework_Yang}, semantic features that exhibit higher relevance to the task are anticipated to make more substantial contributions to the successful execution of intelligent tasks. Additionally, research conducted by \cite{gopika2018correlation} and \cite{jiang2018correlation} has substantiated that feature correlations can inform feature selection, and the use of correlation-based feature weighting can lead to improved task performance. This is attributed to the notion that features demonstrating stronger correlations with other features tend to encapsulate a more abundant information content. Consequently, a holistic approach that considers both these aspects concurrently becomes imperative for an accurate assessment of semantic importance.

As for STR, we first input the image to the pre-trained semantic encoder and decoder to obtain the semantics and the task results. We then compute the gradient of the task result (such as objects, class, and actions) \cite{semantic_concept}, $\boldsymbol{p}$ (before Softmax layer), with respect to the $k$-th semantic feature ${{\boldsymbol{A}}^{_k}}$, i.e., ${\frac{{\partial {\boldsymbol{p}}}}{{\partial {\boldsymbol{A}}^k}}}$. These gradients flowing back are global-average-pooled over the width and height dimensions (indexed by $i$ and $j$ respectively) to obtain the semantic task relevance
\begin{eqnarray}\label{weight1}
g_k = \frac{1}{{W \times H \times N}}{\sum\limits_n{\sum\limits_i {\sum\limits_j {\frac{{\partial {p_n}}}{{\partial {\boldsymbol{A}}_{ij}^k}}} }}}
\end{eqnarray}
where $W$ and $H$ are the width and height of ${{\boldsymbol{A}}^{_k}}$, and ${\boldsymbol{A}}_{ij}^k$ is the activation value at the $i$-th row and the $j$-th column of the feature map. $N$ is the dimension of the task result (e.g. $N=10$ for a ten-class classification task), and $p_n$ is the $n$-th element of the task result $\boldsymbol{p}$. During computation of $\omega _k$ while backpropagating gradients with respect to activations, the exact computation amounts to successive matrix products of the weight matrices and the gradient with respect to activation functions till the final convolution layer that the gradients are being propagated to. Hence, this weight $g_k$ represents a partial linearization of the deep network downstream from ${{\boldsymbol{A}}}$, and captures the ``\emph{semantic task relevance}'' of semantic feature $k$ for a specific task\cite{CAM}. It should be noted that STR is solely dependent on the network parameters, and can be implemented offline once the SemCom system is trained. Consequently, these weights can be regarded as shared knowledge and stored in the knowledge base of the sender and receiver, which may vary for different tasks or applications.

The STR module focuses solely on the correlation between semantic information and the intelligent task, assuming that semantic features are independent of each other. However, in reality, semantic features are not completely decoupled, and different semantic features may capture similar semantics. Therefore, it is necessary to consider inter-semantic relevance when measuring semantic importance. There exist several metrics to quantify the relevance between semantic features, such as the Pearson correlation coefficient and cosine similarity. In this paper, we adopt cosine similarity as the metric of choice, which serves as a representative measure without loss of generality. Consequently, the relevance between semantic feature ${{\boldsymbol{A}}^{k}}$ and ${{\boldsymbol{A}}^{j}}$ can be represented as
\begin{eqnarray}\label{cosine}
\rm{sim}\left( {{{\boldsymbol{A}}^k},{{\boldsymbol{A}}^j}} \right) = \frac{{\left\langle {{{\boldsymbol{A}}^k},{{\boldsymbol{A}}^j}} \right\rangle }}{{|{{\boldsymbol{A}}^k}||{{\boldsymbol{A}}^j}|}}
\end{eqnarray}
where $\left\langle \cdot, \cdot\right\rangle$ is inner product.

The global correlation of a semantic feature is computed as the average correlation between that feature and all other semantic features. Thus, the global relevance of semantic feature ${{\boldsymbol{A}}^{k}}$ can be mathematically represented as follows
\begin{eqnarray}\label{weight2}
{v_k} = \frac{1}{{C - 1}}\sum\limits_{j = 1,j \ne k}^C {\rm{sim}\left( {{A^k},{A^j}} \right)}.
\end{eqnarray}

Based on (\ref{weight1}) and (\ref{weight2}), the final semantic importance can be computed as
\begin{eqnarray}\label{weight}
{\omega _k} = {g_k} \times {v_k}.
\end{eqnarray}

Based on the proposed semantic importance evaluation method in this section, the next section will present the proposed algorithm for solving the semantic importance-aware sub-carrier allocation and bit allocation problem based on reinforcement learning.

\vspace{-0.2cm}
\section{Proposed Algorithm}
\label{sec:SA}
In this section, we first propose an efficient sub-carrier allocation policy, leveraging the determined semantic importance. Subsequently, we present a novel RL algorithm called dynamic proximal policy optimization (DPPO), which incorporates a dynamic action space, to address the bit allocation problem based on semantic importance.

\subsection{Sub-carrier Allocation}
By leveraging the correlation between semantic importance weights and CSI feedback in wireless communication, we can optimize the performance of digital SemCom. Specifically, we aim to transmit the important semantic information over more reliable sub-carriers. Recent research in OFDM has demonstrated accurate CSI estimation at the transmitter side through suitable algorithms and feedback mechanisms\cite{MIMO-OFDM}. Consequently, the received sub-channel gains ${\hat{h}_i}$ are assumed to be included in the CSI feedback, providing the transmitter with precise knowledge of the sub-channel state across the entire OFDM system. Sub-channels with higher SNRs possess better sub-channel states, and thus enable more reliable transmission of semantic information. Consequently, an appropriate sub-carrier allocation function is needed to map the most important semantics to the best sub-channels, and so forth. In particular, sub-carriers associated with higher semantic importance weights need to be assigned to more reliable sub-channels. By this way, the sub-carrier allocation problem is transformed into a one-by-one matching problem, which can be efficiently addressed using a greedy algorithm with a time complexity of ${{\mathcal O}}\left( C \right)$, where $C$ is the number of semantics. Based on the low-complexity sub-carrier allocation policy, each semantics is transmitted through different sub-carriers.

\subsection{RL-Based Bit Allocation Algorithm}
In this subsection, we begin by discussing the key components of the DPPO algorithm and the step-by-step process of utilizing it to optimize the quantization bit allocation strategy. Subsequently, we delve into the complexity analysis of the proposed DPPO algorithm.

\subsubsection{Components of the DPPO Algorithm}
In this part, we provide a comprehensive explanation of the components comprising the proposed DPPO algorithm. Once all semantic features have been extracted using the semantic encoder, the quantization process is formulated as a Markov Decision Process (MDP), where each semantic feature undergoes sequential quantization and dequantization based on the allocated bits in index order, ranging from $i = 1$ to $i = C$. The DPPO algorithm consists of four essential components: a) action space, b) state space, c) policy, and d) reward function. These components are outlined as follows:

a) \emph{Action Space}: We define the action of the agent as the number of allocated quantization bits for all users. Since the total number of quantization bits available for the $i$-th step depends on the number of bits allocated to the previous $i-1$ steps, the action space at $i$-th step should be dynamically adjusted accordingly. Therefore, a dynamic action space $\mathcal{A}^i$ is designed as a set of available bits ${a^{(i)}} \in \left[ {1, 2, ..., B_i} \right]$ assigned at the $i$-th step, where $B_i = B - \sum\limits_{i = 1}^{i - 1} {{b_i}}$ is the total number of bits available at the $i$-th step. In the initialization step, we allocate the bits evenly among all the quantizers. It is important to note that higher quantization bits correspond to finer details in the quantization process.

b) \emph{State Space}: The state in $i$-th step includes the semantic feature $\boldsymbol{a}_i$ and the semantic importance $\boldsymbol{\omega}$, which can be denoted as $s^{(i)} = \{\boldsymbol{a}_i, \boldsymbol{\omega}\}$. The state space $\mathcal{S}$ is a continuous space whose size depends on the number of semantic features.

c) \emph{Policy}: The behavior of the agent is defined as a policy, which determines the probability of selecting each action given a particular state, denoted as $\pi \left( {{a^{(i)}}\left| {s^{(i)}} \right.} \right)$. Specifically, in the proposed DPPO algorithm, the policy $\pi$ is implemented via a deep neural network with parameters $\boldsymbol{\theta}$. The neural network takes the state as its input and produces the action as its output, establishing the relationship between the state and the action probabilities.

d) \emph{Reward}: The intermediate reward $r^{i}$ at the $i$-th step is obtained based on the agent $\pi$ taking action ${a^{(i)}}$ at state ${s^{(i)}}$ by evaluating the semantic distortion and the task performance, i.e. $r^{i} \left( {a^{(i)}\left| {s^{(i)}} \right.} \right) = L_0 + L(\boldsymbol{b}^{(i)}) - {\beta}d(\boldsymbol{b}^{(i)})$. The constant term $L_0$ ensures that the reward is greater than zero, and $\boldsymbol{b}^{(i)}$ represents the bit allocation vector at the $i$-th step. If the actor outputs an action that exceeds the available bits at the $i$-th step, a negative reward is given as a penalty. Since the reward function of each step in the proposed DPPO algorithm is equivalent to the objective function of the problem (\ref{Q1}), and DPPO aims to maximize the reward, the proposed DPPO algorithm can solve the optimization problem (\ref{Q1}).

\begin{figure*}[t]
	\begin{center}
		\includegraphics[width=0.8\linewidth]{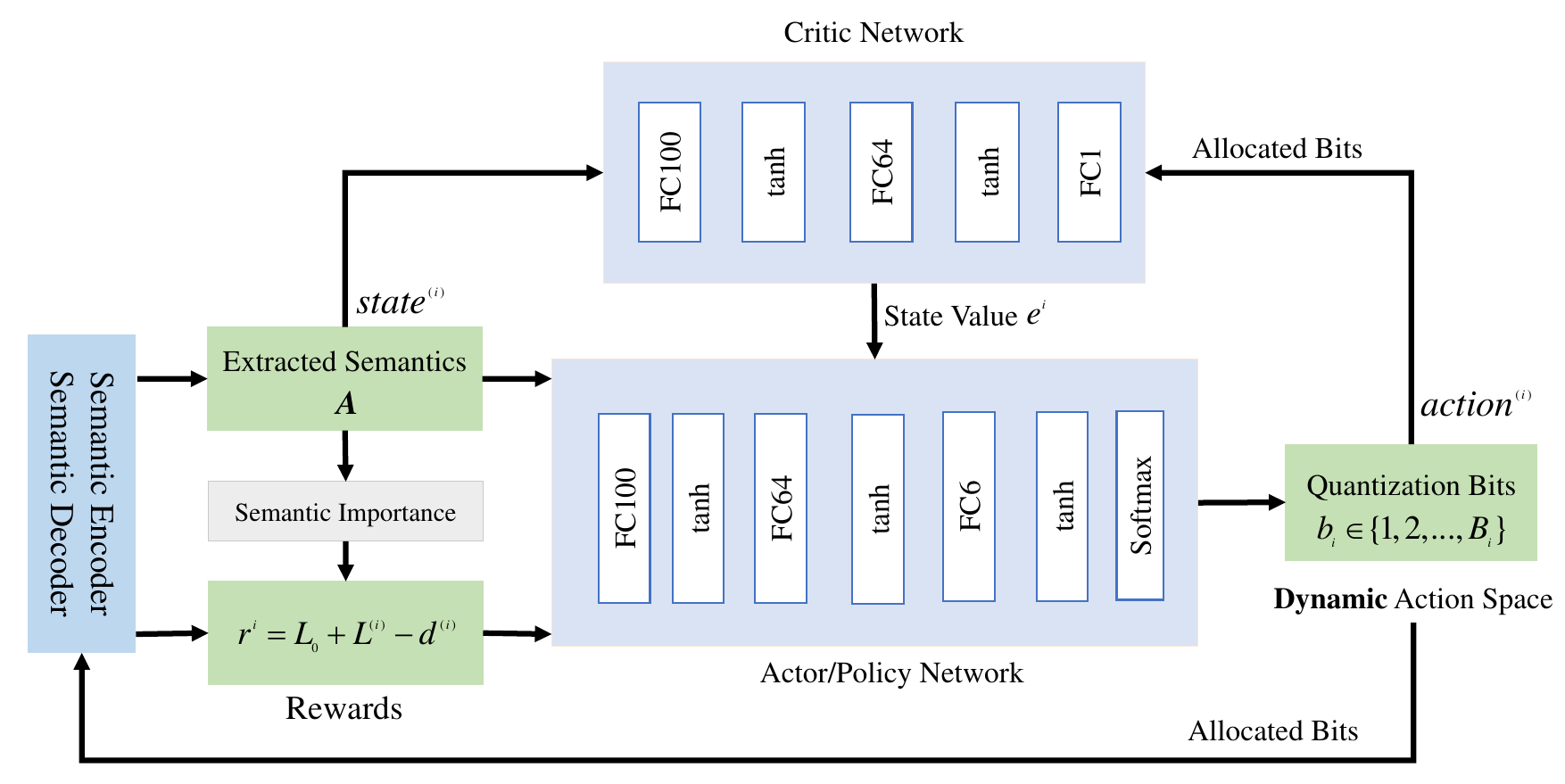}
	\end{center}
	\caption{The architecture of the DPPO-based quantization bit allocation model.}
	\label{fig:RL_structure}
\end{figure*}

\subsubsection{DPPO Algorithm for Bit Allocation}
As illustrated in Fig. \ref{fig:RL_structure}, the architecture of DPPO mainly consists of an Actor network and a Critic network. The actor network processes the input state ${s^{(i)}}$ and consists of three consecutive fully connected (FC) layers. Each FC layer is followed by the Tanh activation function, except for the final layer, which employs a Softmax function to generate a predicted probability distribution over the possible bit allocations. The selected action $a^{(i)}$ is subsequently fed into the critic network, which also consists of three FC layers. The first two layers of the critic network employ the Tanh activation function, while the final layer outputs a state value $e^i$. This state value serves as an estimation of the quality or utility of the selected action at the $i$-th state. Overall, the actor network and critic network work together to determine the action selection process and evaluate the quality of the chosen action, respectively, in order to optimize the bit allocation policy based on the semantic importance.

We then introduce the detailed procedure of the DPPO algorithm. First, the current state ${^{(i)}}$ is fed into the agent to learn the policy $\pi$ that accordingly produces an action $a^{(i)}$ for the $i$-th quantizer. Based on the allocated bits, the semantic features are quantized and transmitted via the wireless channel. Next, the channel decoder takes the received bitstream and the number of allocated bits as input to recover the semantics, which is omitted in Fig. \ref{fig:RL_structure}. Then, the recovered semantics is fed into the semantic decoder to implement the intelligent task and obtain the results. Based on the obtained results, the reward can be obtained, which serves as a guidance signal for updating the policy network.

The whole network can be trained offline, which can be divided into two phases: trajectory sampling and parameter updating. In the trajectory sampling phase, an image is input to the semantic encoder to obtain the corresponding semantics, which is fed into the policy network to obtain a complete trajectory $({{s^{(1)}}}, {a^{(1)}}, r^1, e^1, {{s^{(2)}}}, {a^{(2)}}, r^2, e^2, ..., {{s^{(C)}}}, {a^{(C)}}, r^C, e^C)$. During the training process, the agent is driven to maximize the cumulated reward in the trajectory, which is the optimal goal of the MDP. Denote $\eta$ as the discounted factor, the discounted cumulated reward at timestep $t$ can be expressed as
\begin{eqnarray}\label{cumulated_reward}
R^t = \sum\limits_{k = 1}^t {{\eta ^{t - 1}}{r^t}}.
\end{eqnarray}
Hence, we can obtain the cumulated reward set $\boldsymbol{R} = \{R^1, R^2, ..., R^C\}$ and the state value set $\boldsymbol{e} = \{e^1, e^2, ..., e^C\}$. 

During the parameter updating phase, we first compute the estimator of the advantage function at timestep $t$, denoted as $m^t$, which can be obtained via $m^t = R^t - e^t$. The estimator represents the advantage of the action ${a^{(t)}}$ over the expectation of the action space. Next, we have to obtain the probability ratio between new and old policies, which can be computed by 
\begin{eqnarray}\label{}
{p^t}\left( \theta  \right) = \frac{{{\pi _\theta }\left( {{a^t}\left| {{s^t}} \right.} \right)}}{{{\pi _{{\theta _{old}}}}\left( {{a^t}\left| {{s^t}} \right.} \right)}}, 
\end{eqnarray}
where ${\theta _{old}}$ is the vector of policy parameters before the update. Then, the main objective is \cite{PPO}
\begin{eqnarray}\label{}
{L_1}(\theta ) = \mathbb{E}\left[ {\min \left( {{p^t}(\theta ) {m^t},{\rm{clip}}\left( {{p^t}(\theta ),1 - \varepsilon ,1 + \varepsilon } \right){m^t}} \right)} \right],
\end{eqnarray}
where $\varepsilon$ is a hyper-parameter and $\mathbb{E}$ is the expectation operator. $\rm{clip}(\cdot)$ is a clipping function to modify the surrogate objective by clipping the probability ratio\cite{PPO}. The clipping function limits the range of policy updates to prevent overfitting caused by excessive policy updates.

In line with \cite{PPO}, in addition to the ${L_1}(\theta)$ term, the final loss function should also include a value function error term and an entropy bouns to ensure sufficient exploration, which can be represented as
\begin{eqnarray}\label{loss}
O\left( \theta  \right) = {\mathbb{E}_t}\left[ { - {L_{1t}}(\theta ) + {c_1} {{\left( {{e^t} - {R^t}} \right)}^2} - {c_2}S\left[ {{\pi _\theta }} \right]\left( {{s^t}} \right)} \right],
\end{eqnarray}
where coefficients $c_1$ and $c_2$ are hyper-parameters that control the relative importance of each term in the overall loss function. They can be adjusted to balance the trade-off between policy optimization, value function accuracy, and exploration. The second term represents the value function error term, which measures the difference between the estimated state value and the actual discounted cumulated reward. Minimizing this term helps the value function accurately estimate the expected rewards. The third term represents the entropy bonus term, which encourages exploration by maximizing the policy entropy. This term helps prevent the policy from becoming too deterministic and encourages it to explore different actions.

The optimal policy $\pi$ can be obtained by performing the gradient descent method on the sampled trajectories. The policy update rule is given as
\begin{eqnarray}\label{update}
{\theta ^{(k)}} \leftarrow {\theta ^{(k - 1)}} - \delta {\nabla _\theta }O(\theta ),
\end{eqnarray}
where $\theta ^{(k)}$ is the parameters of the policy at iteration $k$, and $\delta$ is the learning rate.

By iteratively updating the policy until convergence, the proposed DPPO algorithm can learn the relationship between the bit allocation strategy and the system performance. This allows the algorithm to find a policy for bit allocation that maximizes the SemCom performance. The specific training process of DPPO algorithm is summarized in Algorithm \ref{DPPO}.

\begin{algorithm}[t]
	\normalsize
	\caption{Training Process of Proposed DPPO Algorithm.}
	\begin{algorithmic}[1]
		\STATE \textbf{Input:} Image dataset $\boldsymbol{D}$, training epoch $E_p$, learning rate $\delta$, discounted factor $\eta$, hyper parameter $\varepsilon$, $c_1$, and $c_2$, total number of semantics $C$, total number of bits $B$, the pre-trained semantic encoder and semantic decoder.
        \STATE \textbf{Output:} Policy network parameter $\boldsymbol{\theta}$.
		\REPEAT
		\STATE Sample a batch of image data from dataset $\boldsymbol{D}$.
        \STATE Obtain a batch of semantics based on the pre-trained semantic encoder.
        \STATE Compute the semantic importance of each semantics based on (\ref{weight}).
		\STATE Collect trajectories consisting of $C$ timesteps using policy ${\boldsymbol{\theta } _{old}}$.
		\STATE Compute the cumulated reward of each timestep via (\ref{cumulated_reward}) and the loss function via (\ref{loss}).
        \STATE Update the policy parameter $\boldsymbol{\theta}$ by (\ref{update}).
		\UNTIL {the loss function (\ref{loss}) converges.}
	\end{algorithmic}
	\label{DPPO}
\end{algorithm}

\subsubsection{Complexity Analysis of DPPO}
In this subsection, we analyze the computational complexity of the proposed DPPO algorithm for quantization bit allocation. The complexity of the DPPO algorithm is mainly attributed to two factors: calculating the semantic importance using the proposed method and determining the bit allocation using the trained policy. We first analyze the complexity of the proposed semantic importance evaluation method. The STR module can be calculated offline and stored in the knowledge base, so its complexity does not contribute to the online evaluation process. Therefore, the complexity of semantic importance evaluation mainly depends on the ISR module. From (\ref{cosine}), the complexity of calculating the cosine similarity between semantic feature  ${{\boldsymbol{A}}^{k}}$ and semantic feature ${{\boldsymbol{A}}^{j}}$ is ${{\mathcal O}}\left( D \right)$, where $D$ is the size of the semantic feature. From (\ref{weight2}) and (\ref{weight}), the complexity of the proposed semantic importance evaluation method for all semantics can be given as ${{\mathcal O}}\left( CD \right)$, where $C$ is the number of semantics. Next, we investigate the complexity of using the trained policy for bit allocation, which depends on the size of the policy network parameter $\boldsymbol{\theta}$. The size of the policy network parameter $\boldsymbol{\theta}$ lies in the size of action space $\mathcal{A}$ and state space $\mathcal{S}$. In the proposed DPPO, the action space dynamically changes over timesteps, and the size of action space at timestep $t$ is given by $\left| {{{{\mathcal A}}_t}} \right| = B - \sum\limits_{i = 1}^t {{b_i}}$. On the other hand, the size of the state space $\mathcal{S}$ at timestep $t$ is represented by $D$. Therefore, the complexity of using the trained policy to allocate the quantization bits is ${{\mathcal O}}\left( {\sum\limits_{i = 1}^C {\left( {D\left( {B - \sum\limits_{i = 1}^t {{b_i}} } \right)\prod\limits_{l = 2}^L {{H_l}} } \right)} } \right)$, where $L$ is the number of layers in the deep neural network used to train the policy and $H_l$ is the number of neurons in layer $l$. In consequence, the computational complexity of the proposed DPPO algorithm can be expressed as
\begin{eqnarray}\label{complexity}
{{\mathcal O}}\left( {CD + \sum\limits_{i = 1}^C {\left( {D\left( {B - \sum\limits_{i = 1}^t {{b_i}} } \right)\prod\limits_{l = 2}^L {{H_l}} } \right)} } \right).
\end{eqnarray}

From (\ref{complexity}), given the structure of the DPPO model (i.e. $L$ and $H_l$), the complexity of DPPO algorithm depends on the number of semantics, the size of each semantics, and the overall number of bits.

\section{Simulation Results and Analysis}
\label{sec:simulation}
In this section, we perform a comprehensive series of simulations aimed at validating the efficacy of both the proposed OFDM-based digital SemCom system and the DPPO algorithm.

\subsection{Simulation Setup}
1) \emph{Datasets}: Given our focus on the image classification task, we have employed the STL-10 dataset\cite{stl-Coates} as the primary dataset for the OFDM-bssed digital SemCom and for evaluating the proposed DPPO algorithm. The STL-10 dataset holds a prominent position as a benchmark dataset in the realm of image classification, featuring 10 distinct classes and encompassing a rich assortment of labeled images sourced from the extensive ImageNet database.

2) \emph{OFDM symbol structure}: The OFDM symbol structure encompasses a total of 272 sub-carriers. Within a single OFDM symbol, 16 pilots are evenly distributed across 256 data subcarriers. To mitigate inter-symbol interference effectively, the CP length has been set to 72, ensuring it exceeds the delay spread of the channel, as specified in this study.

3) \emph{Channel environment}: To evaluate the robustness of the communication system, we subject it to testing within an untrained channel environment. Specifically, we employ the Stanford University Interim (SUI) channel model, designed to simulate an outdoor multipath channel. In this evaluation, we focus on the fifth scenario of SUI, denoted as SUI-5. SUI-5 comprises three distinct paths positioned at sampling points of delay spread [0, 4, 10] and a corresponding power profile of [0, -5 dB, -10 dB].

4) \emph{Baselines}: We compare the proposed algorithm (labeled as DPPO) with the following conventional methods. a) Evenly allocation method: the quantization bits are evenly allocated to the scalar quantizers, labeled as EAM. b) the ratio-based allocation method: the quantization bits are allocated to the scalar quantizers based on the ratio of corresponding semantic importance, labeled as RBAM. c) random allocation method: the quantization bits are randomly allocated to the scalar quantizers, labeled as RAM. Since all conventional methods do not consider
subcarrier allocation, in order to facilitate the comparison, we first allocated sub-carriers using the proposed algorithm and then allocated bits using the conventional methods. Furthermore, to assess the impact of the ISR module in semantic importance evaluation, we conduct an ablation study by excluding the ISR module, resulting in a simplified version of the DPPO algorithm that considers only semantic-task relevance, denoted as simplified DPPO. Unless specifically stated, all other settings remain unchanged from the original DPPO algorithm.

Top-1 accuracy is used to measure the performance of image classification task. The simulations and experiments are performed by the computer with Ubuntu16.04 + CUDA11.0, and the selected deep learning framework is Pytorch. The number of semantics $C$ is 512 and the total number of bits is [800, 1300]. Other simulation parameters are summarized in Table \ref{tab:experiment}.

\begin{table}[t]
	\normalsize
	\vspace{-0.1cm}
	\centering
	\caption{SIMULATION PARAMETERS.}
	\setlength{\abovecaptionskip}{-0.5cm}
	\begin{tabular}{cm{1.2cm}<{\centering}cc}
		\toprule
		\textbf{Parameter} & \textbf{Value} & \textbf{Parameter} & \textbf{Value} \\
		\hline
		Discount factor, $\eta$ & 0.99 & Hyper-parameter, $\beta$ & 0.5\\
		Training Epoch, $E_p$ & 20 & Hyper-parameter, $\varepsilon$ & 0.25\\
		Optimizer & Adam & Hyper-parameter, $c_1$ & 0.5\\
		Learning rate, $\delta$ & $1\times 10^{-3}$ & Hyper-parameter $c_2$ & 0.01\\
        Drop & 0.3 & Constant, $L_0$ & 10\\
		\toprule
	\end{tabular}
	\vspace{-0.cm}
	\label{tab:experiment}
\end{table}

\subsection{Performance Evaluation of the Proposed Digital SemCom}
In this subsection, we undertake a comparative analysis between our proposed OFDM-based digital SemCom with 64QAM modulation (labeled as DSC) and two baseline systems: analog semantic communication (labeled as ASC) system and a separate source and channel coding (labeled as SSCC) system. The baseline ASC system leverages a cutting-edge image classification-oriented semantic codec, identical to the one utilized in our digital SemCom system. On the other hand, the baseline SSCC system employs JPEG as the image codec and LDPC, as stipulated in the IEEE 802.11n WiFi standard, as the channel code. For modulation, we apply 64QAM to the bit sequences that have undergone JPEG compression and LDPC coding. It is important to note that we consider multipath channel and utilize per-channel MMSE channel estimation, consistent with our proposed system.

\begin{figure}[t]
\centering
\includegraphics[width=1\linewidth]{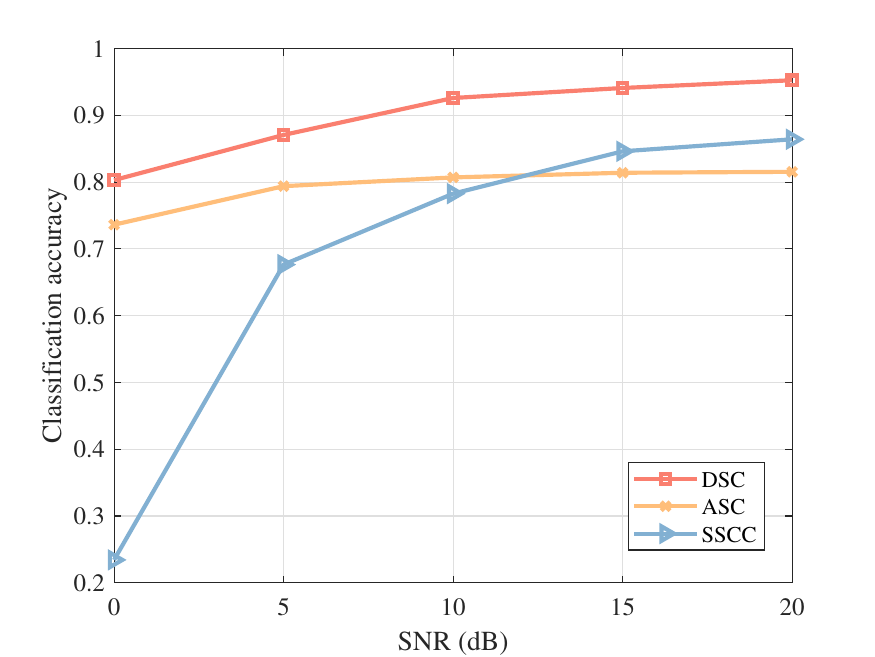}
\caption{Task performance versus channel SNRs for multipath channel model.}
\label{fig:acc_snr}
\end{figure}

Fig. \ref{fig:acc_snr} illustrates the performance comparison between the proposed digital SemCom schemes and the two baselines in the presence of frequency-selective fading channels, with respect to SNR. As depicted in the figure, the accuracy of DSC, ASC and SSCC exhibit an increasing trend as SNR rises, eventually converging towards a specific threshold. This behavior stems from the fact that higher SNR levels correspond to improved communication conditions, resulting in reduced transmission errors and enhanced task performance at the receiver. In addition, the proposed DSC scheme consistently outperforms ASC and SSCC across the entire range of SNR values. Notably, when the SNR is set at 5 dB, the proposed DSC achieves remarkable improvements, resulting in a 9.7\% increase in classification accuracy compared to ASC and a substantial 28.7\% increase compared to SSCC. This disparity arises from the unique strengths of the OFDM-based DSC, which combines the benefits of SemCom and digital communication. Consequently, it can adeptly address channel noise while simultaneously mitigating the impacts of multipath fading. Furthermore, it is noteworthy that SSCC underperforms ASC at low SNR ratios, while the opposite holds true when SNR exceeds 15dB. This is because noise exerts a more substantial impact on communication performance at low signal-to-noise ratios, and ASC demonstrates greater resilience to noise than SSCC. Conversely, in high SNR scenarios, multipath fading assumes a more pronounced role in influencing communication performance, and OFDM-based SSCC proves more adept at combating this effect.

\begin{figure}[t]
\centering
\includegraphics[width=1\linewidth]{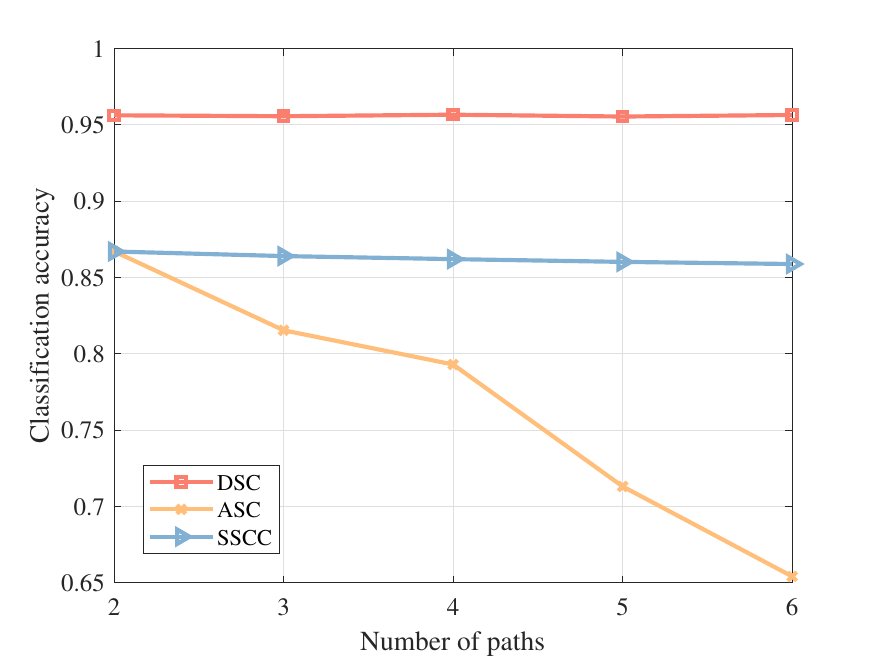}
\caption{Task performance versus the number of paths of multipath channel model.}
\label{fig:acc_paths}
\end{figure}
Fig. \ref{fig:acc_paths} demonstrates the correlation between task performance and the number of paths in the multipath channel model at an SNR of 20 dB. The figure reveals that, as the number of paths increases, the performance of DSC and SSCC remains relatively stable, whereas the performance of ASC exhibits a significant decline. This behavior can be attributed to the fact that a higher number of paths indicates poorer channel conditions and increased interference, thereby negatively impacting the performance of ASC. However, the OFDM-based DSC and SSCC leverage the advantages of existing digital communication technology to effectively mitigate the detrimental effects of these channels, ensuring excellent communication performance even in complex scenarios. This observation also validates the robustness of the proposed method in the presence of multipath channels, which holds great significance for the practical application of SemCom in complex channel environments.

\begin{figure}[t]
\centering
\includegraphics[width=1\linewidth]{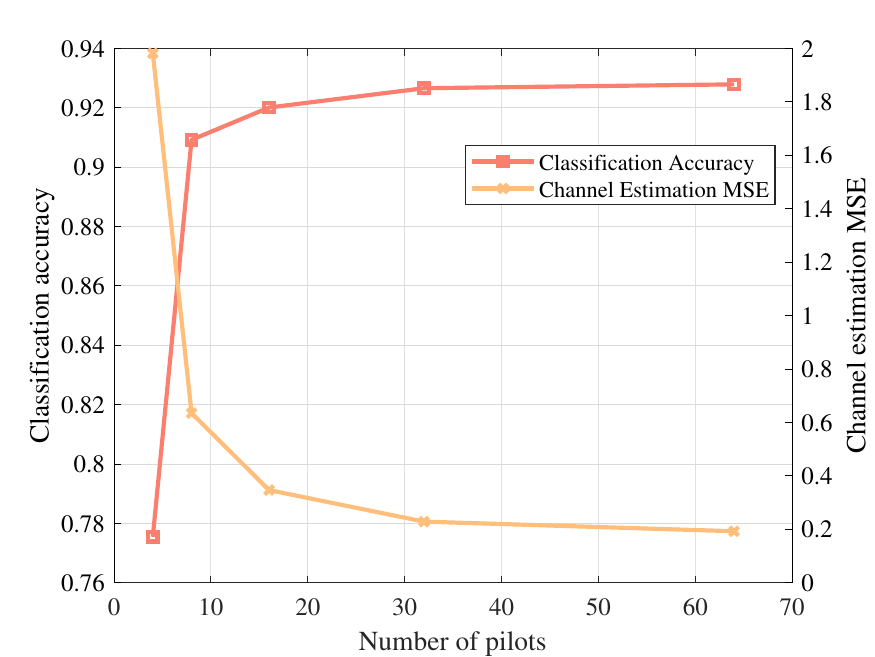}
\caption{Task performance of the proposed method with respect to the number of pilot symbols.}
\label{fig:acc_pilots}
\end{figure}
Fig. \ref{fig:acc_pilots} depicts the relationship between task performance and the number of pilot symbols, as well as the relationship between channel estimation MSE and the number of pilot symbols. As observed, increasing the number of pilot symbols leads to a gradual reduction in channel estimation error and an improvement in intelligent task performance. This outcome can be attributed to the enhanced accuracy of channel state estimation across all subcarriers, enabling better channel equalization at the receiving end of the proposed system to counteract wireless channel interference. However, employing additional pilot symbols results in increased communication overhead, thereby reducing the efficiency of data transmission. Therefore, it is crucial to strike a balance between the two factors. The figure reveals that when the number of pilots is set to 16, the task performance of the proposed method approaches optimality while maintaining a relatively low communication overhead. Consequently, for all subsequent experiments, the number of pilots is fixed at 16.

\subsection{Performance Evaluation of the Proposed DPPO Algorithm}
In this subsection, we first present the convergence curve of the proposed algorithm, and then compare the proposed algorithm with four baselines with respect to task performance and semantic distortion.

\begin{figure}
\centering
\includegraphics[width=1\linewidth]{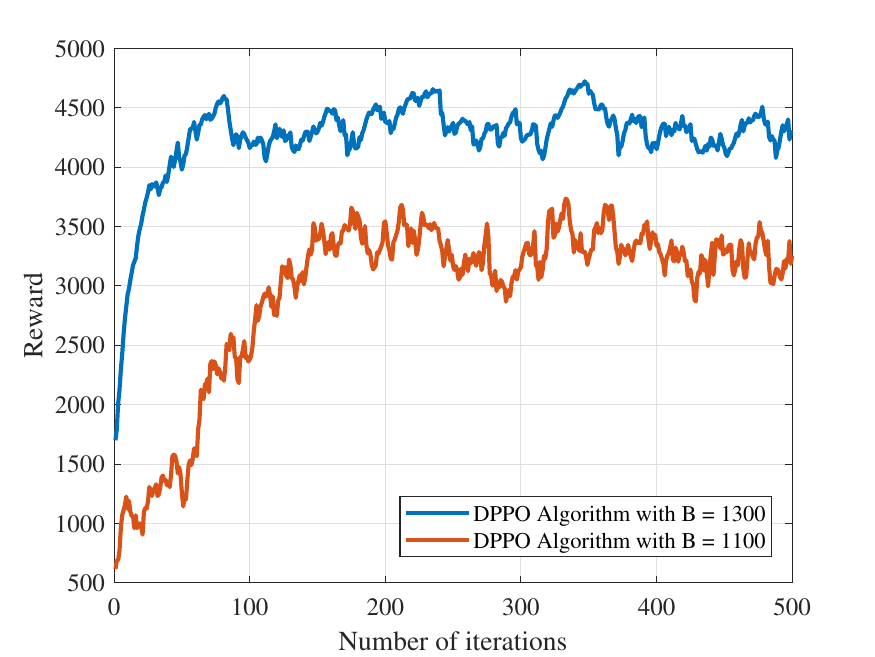}
\caption{The training process of the proposed algorithm.}
\label{fig:convergence}
\end{figure}
Fig. \ref{fig:convergence} illustrates the convergence behavior of the proposed DPPO algorithm within the DSC system, considering two different bit constraints: $B=1300$ bits and $B=1100$ bits. The figure illustrates the training iterations on the x-axis and the corresponding reward on the y-axis. As depicted in the figure, both the DPPO algorithm wiith $B=1300$ and $B=1100$ bits exhibit rapid convergence, reaching a stable performance level in fewer than 200 iterations. This finding emphasizes the fast convergence characteristics of the algorithm, which can be attributed to several key factors. First, the inclusion of a clipping function limits the magnitude of policy changes, enabling more stable updates and preventing extreme policy fluctuations. Additionally, the design of the loss function incorporates action probability distribution entropy, which facilitates effective exploration by the agent, ensuring a robust learning process. The fast convergence of the DPPO algorithm brings notable advantages to the training process of practical DSC systems. By converging quickly, the algorithm reduces the time and energy overhead associated with training, making it more efficient and practical for real-world applications. Furthermore, it is observed that the convergent reward with $B=1300$ bits is higher than the reward with $B=1100$ bits. This difference is due to the higher bit constraint scenario ($B=1300$), allowing for more accurate quantization and transmission of semantics. Consequently, the system with a higher bit constraint achieves better performance in terms of reward.

\begin{figure*}[t]
\centering
\subfigure[Semantic Distortion]{
\includegraphics[width=0.48\linewidth]{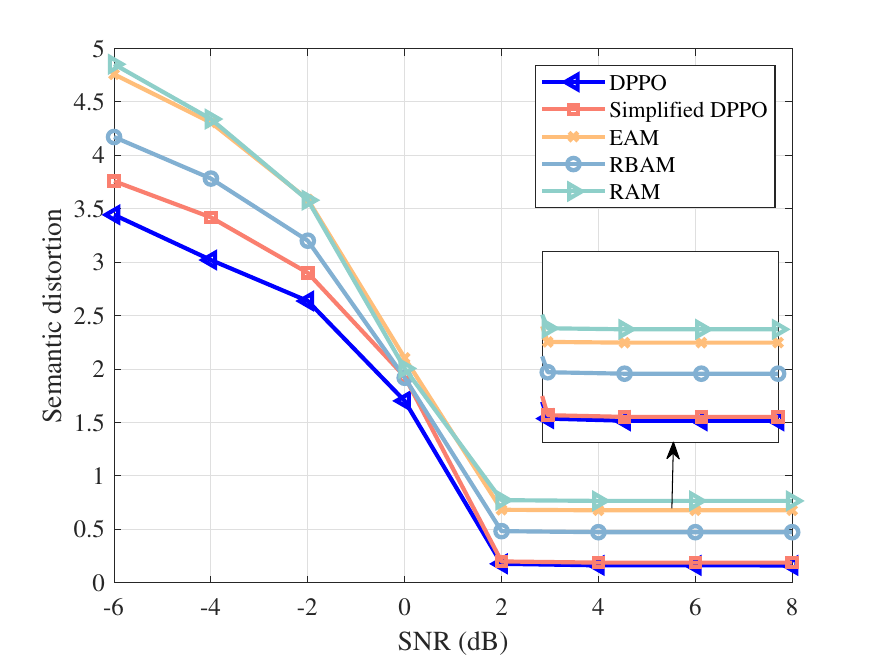}}
\subfigure[Classification Accuracy]{
\includegraphics[width=0.48\linewidth]{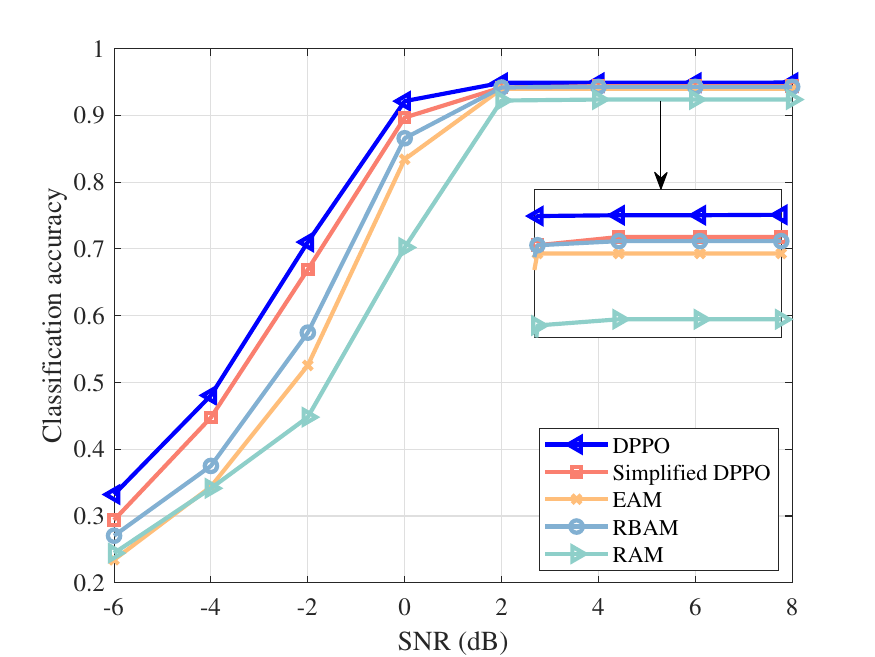}}
\caption{Performance comparison versus channel SNRs under total number of bits equals 1100.}
\label{fig:performance_snr}
\end{figure*}
Fig. \ref{fig:performance_snr} illustrates the comparison of SemCom performance, including semantic distortion and classification accuracy, across different SNRs. It can be observed from the figure that as the SNR increases, there is a decrease in semantic distortion and an increase in classification accuracy, approaching the thresholds. This trend can be attributed to the fact that higher SNR indicates better channel conditions and reduced transmission errors, resulting in improved SemCom performance. Furthermore, the proposed DPPO algorithm and its simplified variant exhibit superior performance compared to the baseline methods in terms of semantic distortion and task performance. Specifically, when the SNR is $-$4dB, the proposed DPPO algorithm improves task performance by 11\%, 14\%, and 14\% compared to the RBAM, EAM, and RAM, respectively. This improvement is attributed to the ability of the proposed algorithm to allocate bits based on semantic importance and the learned policy. RBAM demonstrates improved performance compared to EAM as it can better preserve important semantics to a certain extent. By allocating more bits to important semantics, wireless communication can accurately transmit these crucial semantics, thereby ensuring better SemCom performance. Moreover, the comparison between DPPO and Simplified DPPO reveals that the proposed inter-semantics relevance module plays a vital role in capturing semantic importance and guiding bit allocation. The superior performance of DPPO over Simplified DPPO validates the significance of the inter-semantics relevance module in accurately allocating bits based on semantic importance.

\begin{figure*}[t]
\centering
\subfigure[Semantic Distortion]{
\includegraphics[width=0.48\linewidth]{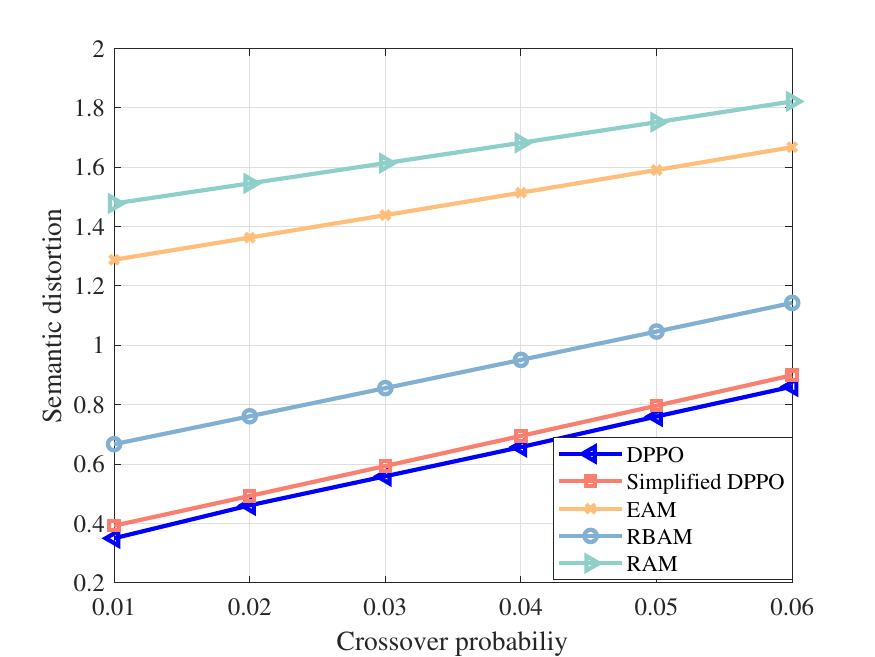}}
\subfigure[Classification Accuracy]{
\includegraphics[width=0.48\linewidth]{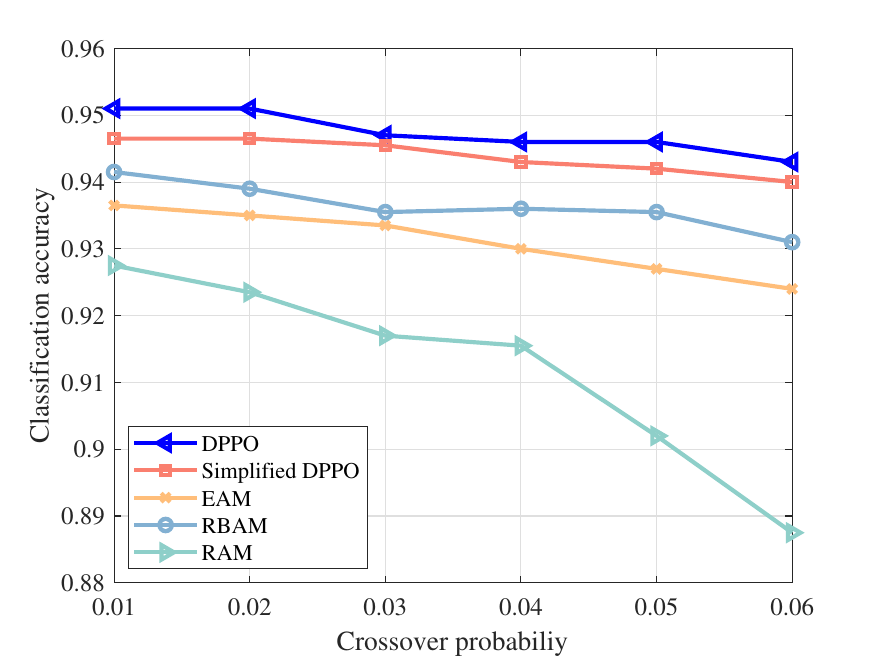}}
\caption{Performance comparison versus crossover probability under BSC channel.}
\label{fig:performance_prob}
\end{figure*}
In addition to multipath channels, Fig. \ref{fig:performance_prob} expands the analysis to include the binary symmetric channel (BSC) \cite{Sun_AIB} and compares the SemCom performance across different crossover probabilities. It is evident from the figure that an increase in crossover probability leads to higher semantic distortion and a decrease in classification accuracy. This behavior can be attributed to the fact that a lower crossover probability indicates better channel conditions and reduces bit errors during wireless transmission. Among all the crossover probabilities considered, the proposed algorithms outperform the baselines in terms of SemCom performance, primarily due to their ability to protect important semantics. Specifically, when the crossover probability is 0.03, the proposed DPPO algorithm achieves a reduction in semantic distortion of 35\%, 61\%, and 66\%, respectively, compared to the three baselines. This indicates that the proposed algorithms excel in maintaining the integrity of important semantics, leading to improved performance. Similar to Fig. \ref{fig:performance_snr}, Fig. \ref{fig:performance_prob} also supports the effectiveness of the ISR module in accurately measuring semantic importance and guiding bit allocation. By leveraging the ISR module, the proposed algorithms are able to allocate bits more effectively, leading to enhanced performance.

\begin{figure*}[t]
\centering
\subfigure[Semantic Distortion]{
\includegraphics[width=0.48\linewidth]{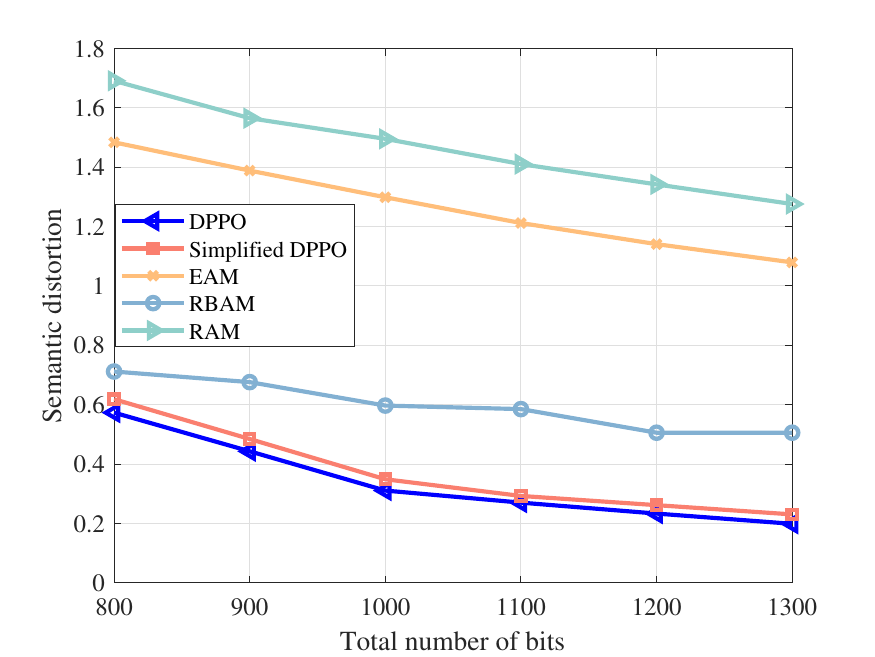}}
\subfigure[Classification Accuracy]{
\includegraphics[width=0.48\linewidth]{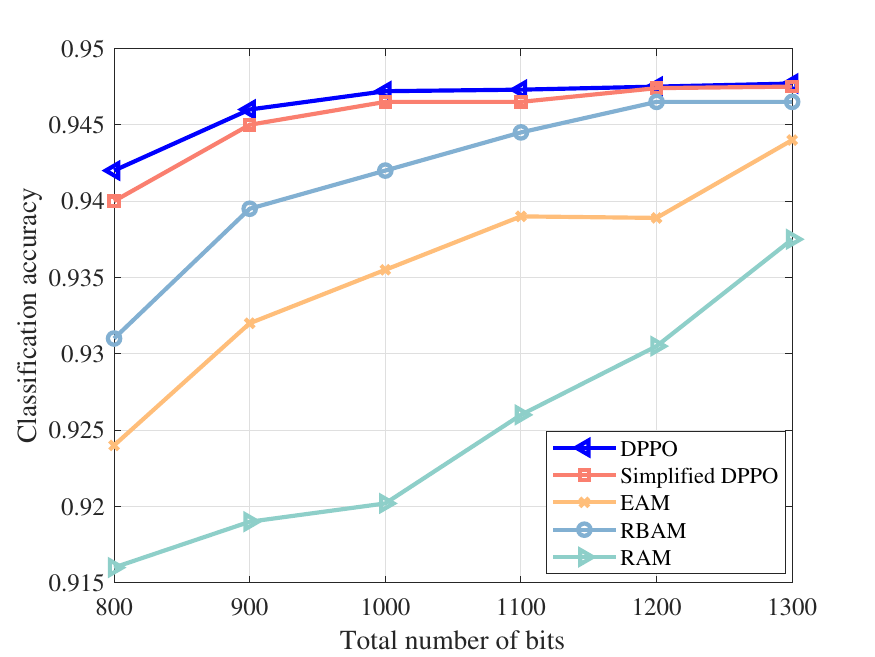}}
\caption{Performance comparison versus the total number of bits.}
\label{fig:performance_bits}
\end{figure*}
Fig. \ref{fig:performance_bits} presents a comparison of SemCom performance across different total numbers of bits. From these figures, it can be observed that semantic distortion decreases and classification accuracy increases as the total number of bits increases. This relationship stems from the fact that a higher total number of bits allows for more accurate transmission of semantics. As more bits are available, the communication system can convey the semantics with greater precision. As the total number of bits increases, the performance gap between the baselines and the proposed algorithm diminishes. This trend occurs because, when the total number of bits is sufficient, all semantics can be accurately transmitted. This finding highlights the effectiveness of the proposed algorithm, particularly when the number of bits is limited. The algorithm demonstrates its ability to optimize bit allocation and improve SemCom performance under constrained conditions. Additionally, it is worth noting that the RAM baseline consistently performs the worst across the various total numbers of bits. This outcome can be attributed to the random bit allocation approach, which fails to ensure the accurate transmission of important semantics. Moreover, there is a possibility of allocating more bits to unimportant semantics, resulting in the inefficient utilization of bit resources. The simulation results affirm the critical role of reasonable bit allocation in enhancing SemCom performance. These findings provide valuable insights for guiding future optimization efforts in SemCom transmission.

\begin{figure*}[t]
\centering
\subfigure[Semantic Importance Visualization]{
\includegraphics[width=0.46\linewidth]{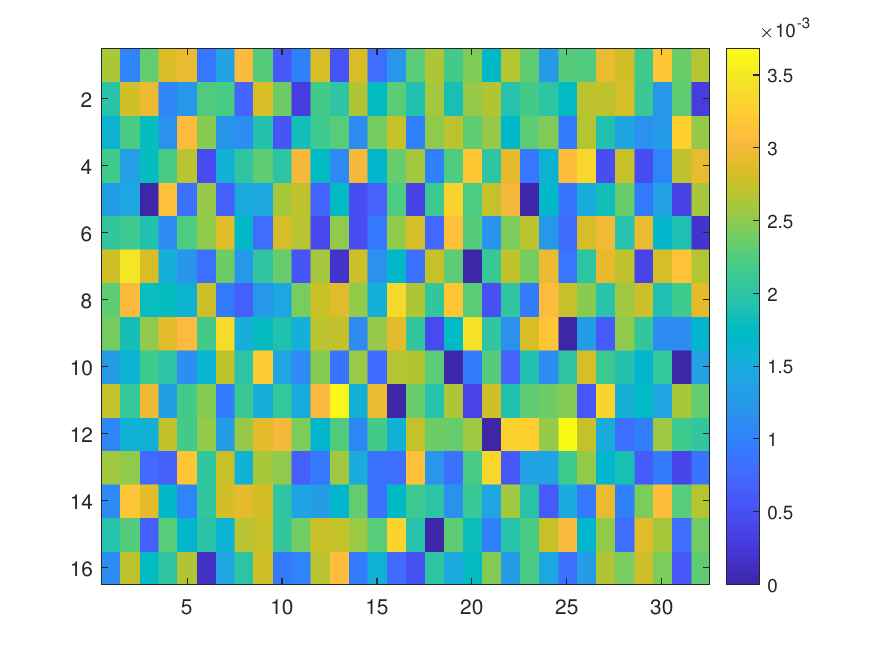}}
\subfigure[Bit Allocation Results]{
\includegraphics[width=0.46\linewidth]{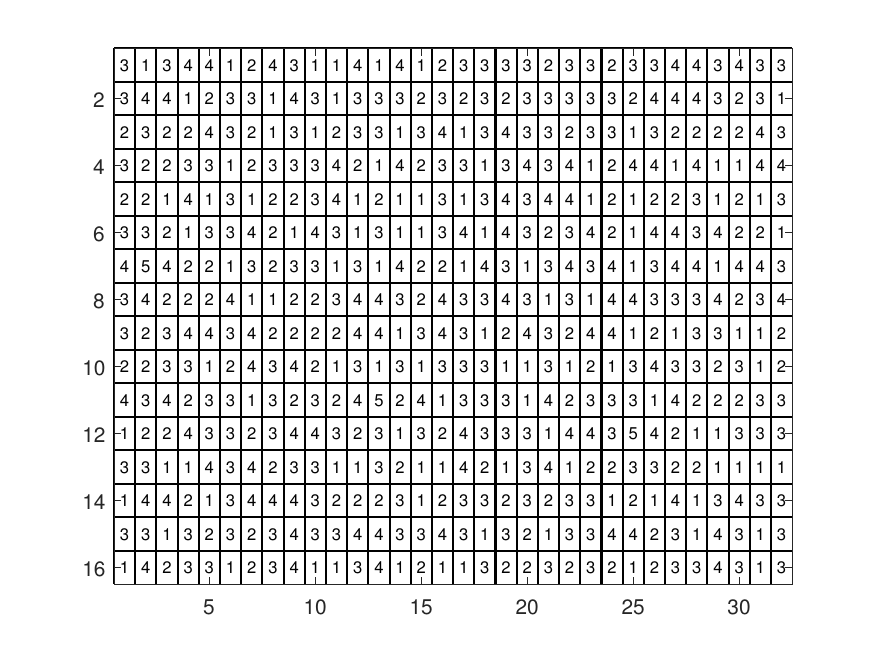}}
\caption{An example of semantic importance-aware bit allocation.}
\label{fig:visualization}
\end{figure*}
Fig. \ref{fig:visualization} showcases an example that demonstrates the importance distribution of extracted semantics and the corresponding bit allocation results. In Fig. \ref{fig:visualization}(a), the semantic importance of the transmitted semantics is visualized, with each semantic represented by a different color according to its importance level. The color spectrum, ranging from blue to yellow, signifies an increasing degree of importance. This example specifically considers an image with 512 semantic feature maps, and the importance values of all the semantics in the image are normalized to have a sum of 1. Moving on to Fig. \ref{fig:visualization}(b), it presents an instance of the resulting bit allocation based on the semantic importance. Notably, the allocation of bits is adapted to the importance of each semantic. More crucial semantics receive a larger number of bits, ensuring their accurate transmission. This allocation strategy proves beneficial, leading to reduced semantic distortion and improved task performance. By prioritizing the allocation of a greater number of bits to important semantics, the proposed approach enhances the fidelity of SemCom. The example depicted in Fig. \ref{fig:visualization} serves as a visual demonstration of the effectiveness of the proposed method in allocating bits based on semantic importance. It highlights the significance of considering semantic importance in the allocation process, resulting in improved overall performance. These findings contribute to advancing the optimization of SemCom systems.

\section{Conclusion}
\label{conclusion}
In this paper, we present an innovative OFDM-based digital SemCom system that seamlessly integrates with existing communication infrastructures while effectively handling complex channels. Within the established system, we have devised an efficient semantic importance measurement method considering both semantics-task relevance and inter-semantics relevance. Based on this, we further propose a low-complexity sub-carrier allocation method and a RL-based bit allocation algorithm, DDPO. Through extensive experiments, we have demonstrated that our digital SemCom method surpasses analog SemCom systems, particularly in challenging multi-path channel scenarios. Furthermore, simulation results have revealed that the DDPO algorithm outperforms various bit allocation baselines, especially in bits-constraint environments. The results of this research indicate the significant potential of our proposed system and algorithms in advancing SemCom technology and achieving improved performance in real-world communication scenarios.

%
\bibliographystyle{IEEEbib}
\nocite{*}\bibliography{stimreference}

\end{document}